\documentclass[
twocolumn,
floatfix,superscriptaddress,a4paper,showkeys,nofootinbib,reprint]{revtex4-1}
\usepackage[colorlinks=true,linktocpage=true,linkcolor=blue,citecolor=blue,allcolors=blue]{hyperref}
\usepackage{epsfig}
\usepackage{latexsym}
\usepackage[utf8]{inputenc}
\usepackage{xspace}
\usepackage{indentfirst}
\usepackage{enumitem}
\usepackage{color}
\usepackage{placeins}
\usepackage[table,xcdraw]{xcolor}
\usepackage{setspace}
\usepackage{lipsum}
\usepackage{multirow}
\usepackage{hyperref}

\usepackage{graphicx}

\usepackage{todonotes}

\usepackage{amsmath}
\usepackage{amssymb}
\usepackage[english]{babel}
\usepackage{url}
\graphicspath{{figs/}}
\newcommand{\const}{\mathop{\rm const}\nolimits}

\newcommand{\eq}[1]{\begin{align} #1 \end{align}}

\newcommand{\sNN}{\sqrt{s_{NN}}}

\newcommand{\be}{\begin{equation}}
\newcommand{\ee}{\end{equation}}

\begin{document}

\title{Chemical freeze-out curve in heavy-ion collisions and the QCD critical point}

\author{Artemiy Lysenko}
\affiliation{Physics Department, Taras Shevchenko National University of Kyiv, 03022 Kyiv, Ukraine}
\affiliation{Bogolyubov Institute for Theoretical Physics, 03680 Kyiv, Ukraine}

\author{Mark I. Gorenstein}
\affiliation{Bogolyubov Institute for Theoretical Physics, 03680 Kyiv, Ukraine}

\author{Roman Poberezhniuk}
\affiliation{Physics Department, University of Houston, Box 351550, Houston, Texas 77204, USA}
\affiliation{Bogolyubov Institute for Theoretical Physics, 03680 Kyiv, Ukraine}

\author{Volodymyr Vovchenko}
\affiliation{Physics Department, University of Houston, Box 351550, Houston, Texas 77204, USA}

\begin{abstract}
    The chemical freeze-out curve in heavy-ion collisions is investigated in the context of a quantum chromodynamics (QCD) critical point~(CP) search at finite baryon densities.
    Taking the hadron resonance gas picture at face value, chemical freeze-out points at a given baryochemical potential provide a lower bound on the possible temperature of the QCD CP.
    We first verify that the freeze-out data in heavy-ion collisions are well described by a constant energy per particle curve, $E/N = \rm const$, under strangeness neutrality conditions~($\mu_S \neq 0$, $\mu_Q \neq 0$).
    We then evaluate the hypothetical lower bound on the freeze-out curve based on this criterion in the absence of strangeness neutrality~($\mu_S = 0$, $\mu_Q = 0$) and confront it with recent predictions on the CP location.
    We find that recent estimates based on Yang-Lee edge singularities from lattice QCD data on coarse lattices~($N_\tau = 6$) place the CP significantly below the freeze-out curve, hinting at the importance of performing continuum extrapolation within this method. Predictions based on functional methods and holography place the CP slightly above the freeze-out curve, indicating that the QCD CP may be located very close to the chemical freeze-out in $A$+$A$ collisions at $\sNN = 3.5$--$6$~GeV.
\end{abstract}

\maketitle

\section{Introduction}

Recently, quantum chromodynamics (QCD) turned 50 years old. Identifying the phases and structure of strongly interacting matter is one of the outstanding issues in modern nuclear physics. A chiral crossover transition at vanishing baryon density and a pseudocritical temperature $T_{pc} \approx 155$~MeV has been well established~\cite{Aoki:2006we,Borsanyi:2010bp,HotQCD:2014kol}. Whether the analytic crossover turns into a first-order phase transition at finite baryon density at a QCD critical point~(CP) remains an open question studied both theoretically, and experimentally through heavy-ion collisions~\cite{Stephanov:1998dy,Stephanov:1999zu,STAR:2020tga,HADES:2020wpc}.

First-principle lattice QCD constraints on the QCD CP come to rely on extrapolations from vanishing and imaginary $\mu_B$. The state-of-the-art results show no indications for the CP at small baryon densities and disfavor its existence at $\mu_B/T \lessapprox 3$~\cite{Borsanyi:2020fev,HotQCD:2018pds,Vovchenko:2017gkg}. Similar conclusion can be drawn from the measurements of proton number cumulants in heavy-ion collisions~\cite{ALICE:2019nbs,STAR:2021iop}, which are consistent with non-critical effects at collision energies $\sNN \gtrapprox 20$~GeV~\cite{Vovchenko:2021kxx}, where $\sNN$ is the collision energy per nucleon pair in center of mass frame.

Theoretical predictions for the QCD CP have also evolved from early days~\cite{Stephanov:2004wx}, now being increasingly constrained to lattice QCD data at $\mu_B = 0$. These are based either on the expected universal behavior of Yang-Lee edge singularities in the vicinity of the CP and the extrapolation of the lattice data to lower temperatures~\cite{Basar:2023nkp,Clarke:2024ugt}, on the analysis of contours of constant entropy density~\cite{Shah:2024img}, or through effective QCD approaches such as Dyson-Schwinger equations~\cite{Gunkel:2021oya,Gao:2020fbl}, functional renormalization group~\cite{Fu:2019hdw}, or holography~\cite{Hippert:2023bel}. These different predictions place the QCD CP in roughly the same ballpark of $T_{CP} \approx 90$--$120$~MeV and $\mu_{B, \ CP} \approx 400$--$650$~MeV.

These results indicate the possibility that the CP can be explored in heavy-ion collisions that probe baryon-rich matter. 
Analysis of hadron production in heavy-ion collisions  performed over many years across many collision energies indicate that a significant degree of chemical equilibration is reached, and that heavy-ion hadron matter is well described by a hadron resonance gas at freeze-out~\cite{Andronic:2017pug}.
Chemical freeze-out curve maps different collision energies onto the $(T,\mu_B)$ phase diagram~\cite{Cleymans:1998fq,Cleymans:2005xv}.
It remains an open question, however, how close these CP estimates are to the freeze-out curve in heavy-ion collisions. 
The majority of the CP estimates have been obtained at vanishing charge and strangeness chemical potentials, $\mu_Q = \mu_S = 0$, while in heavy-ion collisions, these are non-zero due to strangeness neutrality and the initial isospin asymmetry in colliding nuclei. 
Here, we address this issue by reevaluating the chemical freeze-out curve and its hypothetical counterpart under $\mu_Q = \mu_S = 0$ conditions.

First, we show that a constant energy per particle line, $E/N = 0.951$~GeV in the hadron resonance gas~(HRG) model under strangeness neutrality and charge-to-baryon ratio $Q/B = 0.4$ describes well the available chemical freeze-out data in a broad collision energy range $\sNN = 2.4$--$5020$~GeV. Next, we evaluate the same $E/N = 0.951$~GeV line at vanishing charge and strangeness chemical potentials, $\mu_Q = \mu_S = 0$, which would correspond to the hypothetical heavy-ion collisions with the corresponding electric charge and strangeness content. We then compare this line in the $T$-$\mu_B$ plane with the recent CP estimates and discuss how close they are to the freeze-out curve and at which energies.

The paper is organized as follows. In Sec.~\ref{sec:HRG} we briefly revisit the HRG model. In Sec.~\ref{sec:FOline} we discuss the world data of chemical freeze-out points in the $T$-$\mu_B$ and how these can be described by the constant energy per particle criterion. In Sec.~\ref{sec:CPs}, we examine the chemical freeze-out line in the context of QCD critical point search and reevaluate the $E/N = \rm const$ line under $\mu_Q = \mu_S = 0$ conditions. Summary in Sec.~\ref{sec:summary} closes the article.

\begin{figure*}
    \includegraphics[scale=0.5]{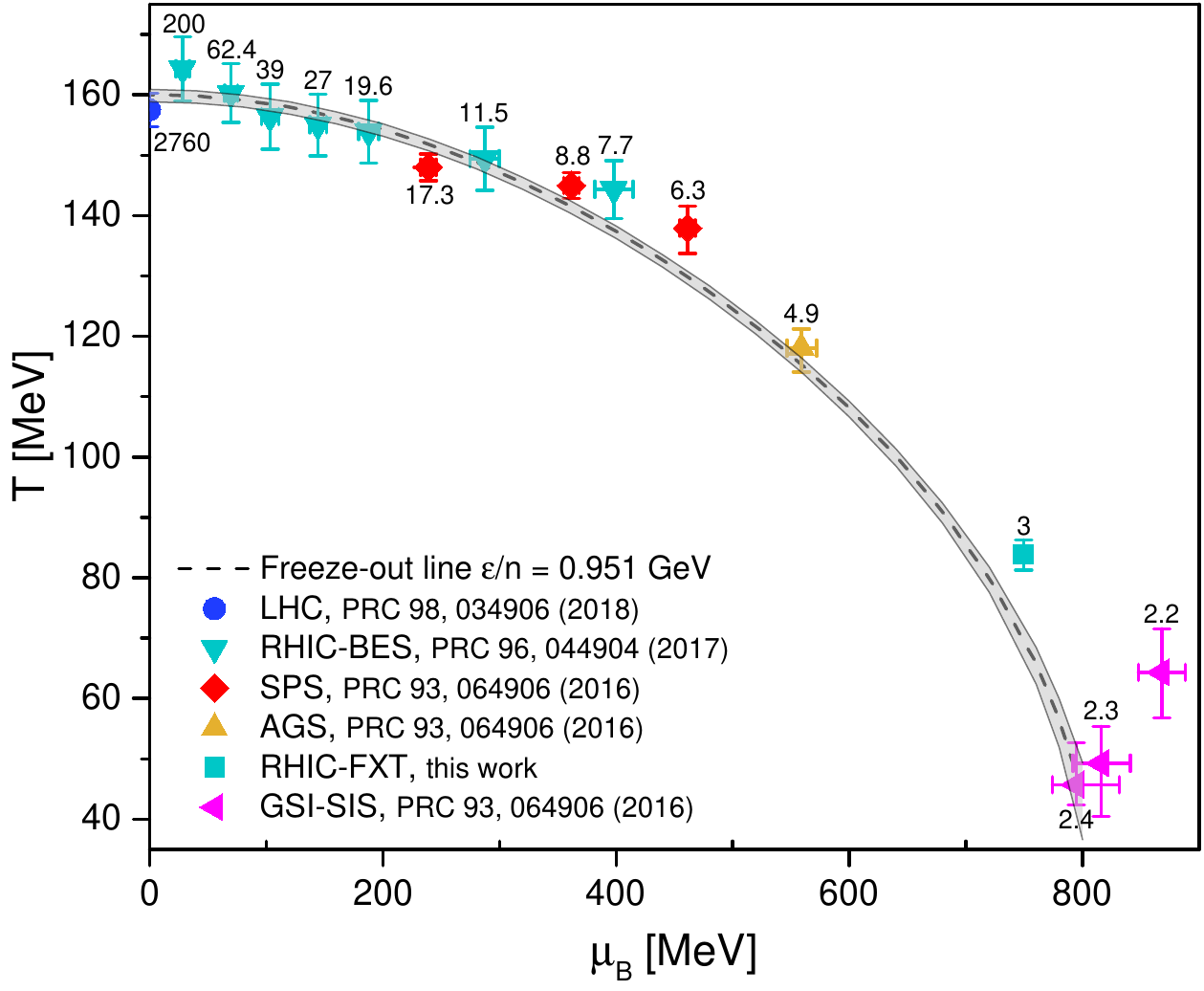}
    \caption{ 
    The symbols depict the various chemical freeze-out data points from the literature~\cite{Vovchenko:2015idt,Vovchenko:2018fmh, STAR:2017sal} and include the 3 GeV data point obtained in this paper~(see text for details). Different symbol styles represent different experiments with the labels indicating $\sNN$~[GeV] for each point. The dashed black line shows the constant energy per particle line $\varepsilon/n = 0.951$ GeV, computed within the ideal HRG model and used in this paper as a parametrization of the freeze-out curve. For clarity, the black dashed line passes through the gray area. The lower limit of the gray area corresponds to the line of constant energy per particle $\varepsilon/n = 0.940$ GeV, and the upper limit to the line $\varepsilon/n = 0.960$ GeV.
    }
    \label{fig:f-o}
\end{figure*}

\section{Hadron resonance gas model}
\label{sec:HRG}
 
We use the hadron resonance gas (HRG) model~\cite{Hagedorn:1965st,Dashen:1969ep,Venugopalan:1992hy} in our calculations. The model describes the equilibrated hadronic system as a multicomponent gas of hadrons and resonance. The essence of this model is described in detail in Refs.~\cite{Cleymans:1992zc,Vovchenko:2014pka}. We perform most of our calculations within the ideal HRG~(Id-HRG) model but also consider its two extensions -- the excluded volume hadron resonance gas (EV-HRG)~\cite{Vovchenko:2017xad} and quantum van der Waals hadron resonance gas (QvdW-HRG)~\cite{Vovchenko:2016rkn}. The calculations are performed in the grand-canonical ensemble.

In the case of the  Id-HRG, the grand canonical pressure, energy density, and particle number densities are calculated as follows:
\eq{\label {eid}
p(T,\mu) & =  \sum_i p_i^{\rm id}(T,\mu_i), \\\nonumber
p_i^{\rm id}(T,\mu_i) & = \frac{d_{i}}{6\pi^{2}} \int dm f_{i}(m) \int_{0}^{\infty} \frac{dk~k^{4}} {  \sqrt{k^{2} + m^{2}}} \\\label{pid}
 &  \times \left[\exp\left(\frac{\sqrt{k^{2} + m^{2}} - \mu_{i}}{T}\right) \pm 1 \right]^{-1}, \\
n_i^{\rm id}(T, \mu_i) & = \left(\frac {\partial p^{\rm id}}{\partial \mu_i}\right)_T, \\
\varepsilon^{\rm id}(T,\mu) & = T \left(\frac{\partial p^{\rm id} }
 {\partial T}\right)_{\mu} 
  + \sum_i \mu_i n_i^{\rm id} 
  -~p^{\rm id}~,
\\
\mu_i & = b_i\mu_B+q_i\mu_Q+s_i\mu_S~,
\label{mui} 
}

\hspace{-0.35cm}where $\mu\equiv (\mu_B,\mu_Q,\mu_S)$ and $\varepsilon^{\rm id}$ is the energy density, $n^{\rm id}$ is the particle number density, $\textit{d}_{i}$ is the number of internal degrees of freedom, $\mu_i$ is the chemical potential of $i$th type of hadrons. Three fundamental chemical potentials $\mu_B$, $\mu_Q$, and $\mu_S$ regulate the three conserved charges in strong interactions, and  $b_i$, $q_i$, and $s_i$ are, respectively, the baryonic, electric, and strange charges of hadrons of the $i$th type. In Eq. (\ref{pid}) "-1" is chosen for bosons, and "+1" - for fermions.

Resonances are also included in the HRG model. The resonance widths are  taken into account as the integrals over their masses $\int dm f_{i}(m)$  in Eq.~(\ref{eid}). They are calculated according to the energy-dependent Breit-Wigner prescription~\cite{Vovchenko:2018fmh}.

We also consider an extension of the Id-HRG model (see the Appendix) that takes into account the repulsive and attractive interactions among (anti)baryons. This is done within the QvdW-HRG model (Ref.~\cite{Vovchenko:2016rkn}). Calculations are performed using the Thermal-FIST (Ref.~\cite{Vovchenko:2019pjl}) software, which is freely available.

\section{Chemical freeze-out line and constant energy per particle}
\label{sec:FOline}

Chemical freeze-out (F-O) is a stage of nucleus-nucleus collisions where the chemical composition of (primordial) hadrons becomes fixed, with subsequent changes in hadron abundances occurring only through resonance decays. At each collision energy, chemical F-O corresponds to a pair of $T$-$\mu_B$ values, with primordial hadron abundances computed in the framework of the HRG model. This simple model has proven to be very successful in describing the measured yields of many different hadron species in heavy-ion collisions across a broad collision energy range~\cite{Cleymans:1999st,Becattini:2000jw,Becattini:2003wp,STAR:2017sal,Andronic:2017pug}. Figure~\ref{fig:f-o} depicts various F-O points from different experiments, including STAR-BES~\cite{STAR:2017sal}, NA49~\cite{NA49:2002pzu,NA49:2006gaj,NA49:2007stj,Vovchenko:2015idt}, E-802~\cite{E802:1999hit,E-802:1999ewk,Vovchenko:2015idt}, and GSI-SIS~\cite{Averbeck:2000sn}. To fill the gap at high $\mu_B$, we additionally performed fits to the recent data for $0$--$10\%$ central Au-Au collisions at $\sNN = 3$~GeV from the STAR Collaboration~\cite{STAR:2023uxk,STAR:2024znc} using Thermal-FIST and the Id-HRG model in the strangeness-canonical ensemble. Namely, we fitted the $4\pi$ yields of primordial protons, light nuclei ($d, {}^{3}\text{H}, {}^{3}\text{He}, {}^{4}\text{He}$), $\langle {\rm N}_{part} \rangle$, and the yields of $\Lambda, K^{0}_S, K^-, \phi, \Xi^-$ normalized by $\langle {\rm N}_{part} \rangle$. This point is shown in Fig.~\ref{fig:f-o} by the turquoise square.

By varying the collision energy, one obtains the so-called chemical F-O line in the $T$-$\mu_B$ plane. There have been different chemical F-O criteria proposed in the literature to describe the F-O line~\cite{Cleymans:1998fq,Cleymans:1999st,Cleymans:2005xv}. Perhaps the most robust criterion corresponds to the constant energy per particle of $E/N \equiv \varepsilon / n \approx 0.9$--$1.0$~GeV~\cite{Cleymans:2005xv} at a F-O, where $n = \sum_i n_i$. One motivation for this criterion is that below a certain energy per particle, the inelastic reactions become too weak to maintain chemical equilibrium in a rapidly expanding system created in heavy-ion collisions. 
The validity of such criterion has also been verified with microscopic hadronic transport simulations~\cite{Reichert:2020yhx}.

We use the $\varepsilon / n = \rm const$ criterion to estimate the chemical freeze-out line.
Figure~\ref{fig:f-o} depicts the line
\eq{ \label{umova1}
    \frac{\varepsilon}{n} = 0.951^{+0.009}_{-0.011} \ \rm GeV 
}
calculated in the Id-HRG model. 
The central $\varepsilon/n$ value corresponds to the chemical freeze-out temperature $T_{ch} = 160$~MeV at $\mu_B=0$, which is in line with thermal fits at CERN Large Hadron Collider (LHC) energies performed within the same model using Thermal-FIST code from Ref.~\cite{Vovchenko:2019pjl}~(blue point in Fig.~\ref{fig:f-o}).
The upper and lower errors have been obtained by requiring the $\varepsilon / n = \rm const$ to graze the lower error bar of $\sNN = 8.8$~GeV and $\sNN = 4.9$~GeV F-O points, respectively.
These requirements are in place to interpret the $\varepsilon/n = \rm const$ line as a lower limit on the F-O temperature at $\mu_B > 400$~MeV, where the CP may be located.
\begin{figure*}
    \includegraphics[width=.49\textwidth]{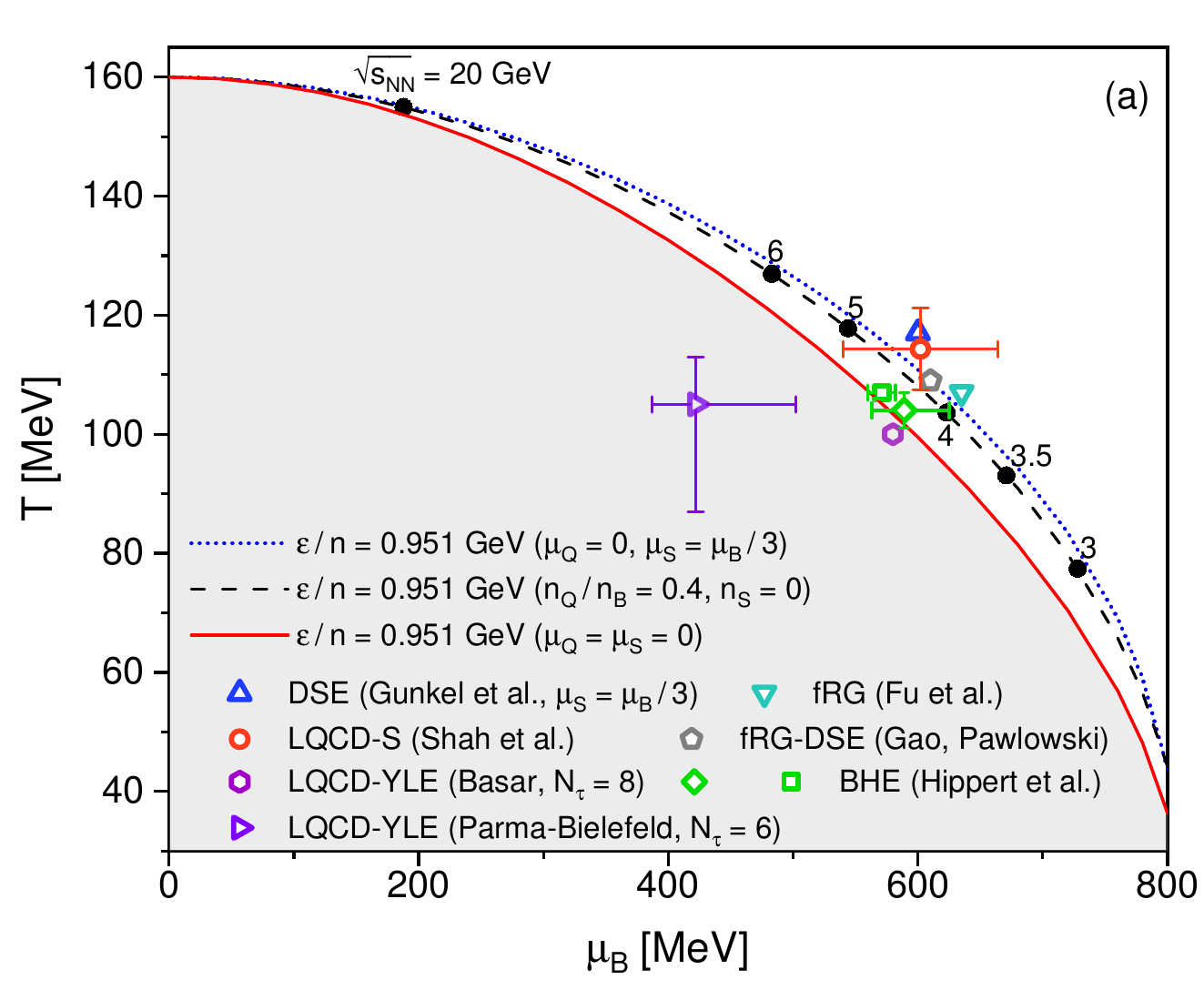}
    \includegraphics[width=.49\textwidth]{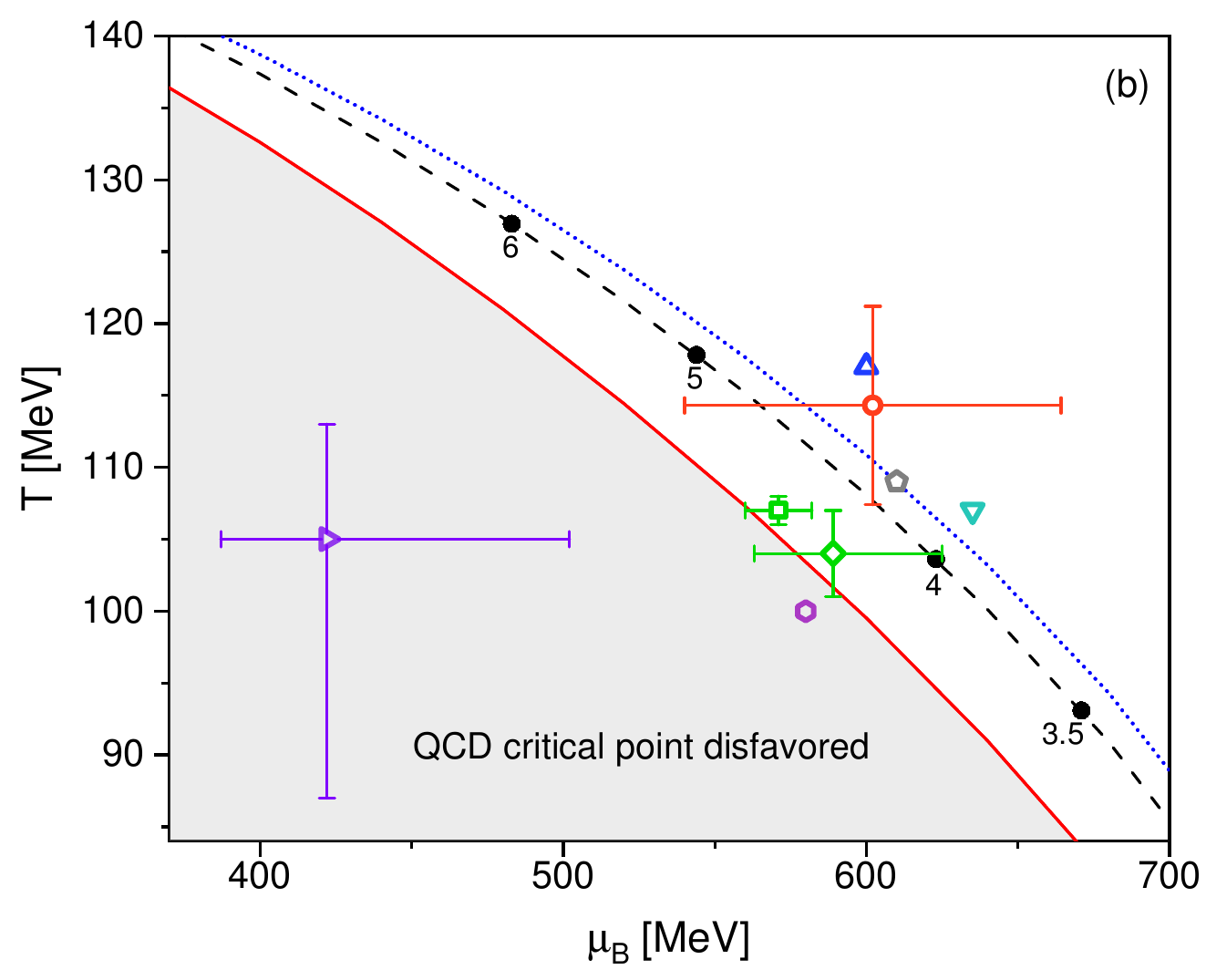}
    \caption{\textbf{(a)} The lines represent the chemical freeze-out curves computed through constant energy per particle criterion $\varepsilon/n = 0.951 \ \rm GeV$ in the ideal HRG model under (dashed black line) heavy-ion collision conditions~($n_Q/n_B = 0.4$ and $n_S = 0$), (solid red line) vanishing charge and strangeness chemical potentials~($\mu_Q = \mu_S = 0$), and (dotted blue line) under the conditions $\mu_Q=0$ and $\mu_S=\mu_B/3$. The shaded area beneath the red line indicates the region of the phase diagram where the QCD critical point is disfavored by heavy-ion data. The black points depict the different $\sNN$ values in GeV. 
    Various critical point predictions from the literature for $\mu_S = 0$~\cite{Fu:2019hdw,Shah:2024img,Hippert:2023bel,Basar:2023nkp,Gao:2020fbl,Clarke:2024ugt} and $\mu_S = \mu_B/3$~\cite{Gunkel:2021oya} are shown by the colored symbols. 
    \textbf{(b)} Zoomed-in view of the region $\mu_{B} \in [370$--$700]$ MeV. 
    }
    \label{fig:frzCP}
\end{figure*}

The line shown in Fig.~\ref{fig:f-o} is calculated taking into account conservation of strangeness $S = 0$ and electric charge $Q/B = 0.4$ (the second condition corresponds to proton-to-nucleon ratio in heavy nuclei). 
By solving these two equations, one determines the values of the electric and strangeness chemical potentials $\mu_Q$ and $\mu_S$ as a function of the baryochemical potential $\mu_B$ and the temperature $T$. 
For this reason, $\mu_{Q}$ and $\mu_{S}$ are not independent variables, and the construction of the  F-O line reduces to solving a system of transcendental equations
\eq{ 
\frac{\varepsilon(T, \mu_{B}, \mu_Q,\mu_S)}{n(T, \mu_{B}, \mu_Q,\mu_S)} & = \const, \\ 
\frac{n_Q(T, \mu_{B}, \mu_Q,\mu_S)}{n_B(T, \mu_{B}, \mu_Q,\mu_S)} & = 0.4, \\ 
n_S(T, \mu_{B}, \mu_Q,\mu_S) & = 0
}
with respect to $T$, $\mu_Q$, and $\mu_S$ at each fixed value of $\mu_B$. One can see from Fig.~\ref{fig:f-o} that the line $\varepsilon(T, \mu_{B}) / n(T, \mu_{B}) = 0.951$~GeV describes fairly well the depicted F-O points, especially at $\mu_B \lessapprox 300$~MeV. At high $\mu_B$, it tends to provide a lower limit on the chemical F-O temperature. 
In fact, it is our main motivation here to obtain a lower temperature limit in the baryon-dense regime for the freeze-out line, which will be relevant for the discussion of the QCD critical point in the following section. As one can see from Fig.~\ref{fig:f-o}, the line in Eq.~\eqref{umova1} captures the lowest possible chemical freeze-out temperature value at $\mu_B > 400$~MeV, thus we take it to define the lower limit on $T_{ch}$ for $\mu_B > 400$~MeV.
In the Appendix (see Fig. \ref{fig:porivHRG}), we show the same calculation within EV-HRG and QvdW-HRG models, which place the F-O line slightly higher at nonzero $\mu_B$. 
We note that the F-O curve may be affected significantly by other choices of repulsive interactions, such as bag model scaling~\cite{Vovchenko:2015cbk,Vovchenko:2016ebv}, but these always tend to increase the F-O temperature. Therefore, the black line in Fig.~\ref{fig:f-o} can indeed be regarded as the lower limit on the F-O curve in heavy-ion collisions.

\section{Chemical freeze-out line and QCD critical point}
\label{sec:CPs}

Taking a picture of an equilibrated hadron resonance gas at chemical F-O at face value; the F-O line provides a lower bound in temperature at fixed $\mu_B$ for the possible location of the QCD CP. 
Decreasing the temperature at a fixed $\mu_B$ leads to smaller hadron number densities. Therefore, the significant effects of deconfinement and chiral symmetry restoration necessary for the CP can only be observed at higher temperatures.

It is instructive to put the lower bound from the F-O curve in the context of recent estimates of the CP location. 
These estimates come from lattice QCD analyses of Yang-Lee edge singularities~\cite{Basar:2023nkp,Clarke:2024ugt}, entropy density contours~\cite{Shah:2024img}, effective QCD approaches, including Dyson-Schwinger equations (DSE)~\cite{Gunkel:2021oya,Gao:2020fbl}, functional renormalization group~(fRG)~\cite{Fu:2019hdw}, and black-hole engineering~(BHE)~\cite{Hippert:2023bel}. 
The corresponding estimates are depicted by symbols in Fig.~\ref{fig:frzCP}. 
One can observe a contradiction of sorts for some of the estimates, namely that they lie below the F-O curve~(dashed black line in Fig.~\ref{fig:frzCP}), where the existence of the CP should be excluded. 
Of course, there is some uncertainty in the F-O curve at large $\mu_B$, but, as discussed in the previous section and visible in Fig.~\ref{fig:f-o}, the black dashed curve generally corresponds to the lower end of the error bar in temperatures.

However, it is important to keep in mind that the F-O curve is obtained under conditions of strangeness neutrality and a charge-to-baryon ratio of $0.4$, which corresponds to significantly non-zero values of $\mu_S$ and $\mu_Q$, roughly $\mu_S \approx \mu_B / 3$ and $\mu_Q \approx -\mu_B/30$. On the other hand, the majority of the theoretical estimates of the CP location in $T$-$\mu_B$ plane are obtained for $\mu_S = \mu_Q = 0$. Resolving this difference in strangeness and charge conservation conditions is necessary to make further conclusions, especially given that most of the CP estimates are only slightly below the chemical F-O curve. Either the CP estimates should be provided under $n_S = 0$, $n_Q/n_B = 0.4$ conditions, or modifications to the F-O curve corresponding to $\mu_S = \mu_Q = 0$ should be evaluated.

Here, we pursue the latter approach and calculate the hypothetical chemical F-O curve under $\mu_S = \mu_Q = 0$ conditions. We do this by calculating the same constant energy per particle line $\varepsilon(T,\mu_B,\mu_Q=0,\mu_S=0)/n(T,\mu_B,\mu_Q=0,\mu_S=0) = 0.951$~GeV in the Id-HRG model. The result is shown by the red line in Fig.~\ref{fig:frzCP}. One can see that the red line lies systematically below the black curve at finite $\mu_B$. Although this downward shift is not dramatic, it is significant for comparisons of CP estimates with the F-O curve. 

In addition, we also consider the case $\mu_S = \mu_B / 3$ and $\mu_Q = 0$, which corresponds to vanishing strange quark chemical potential, $\mu_s = 0$, and describes
strangeness neutral isospin-symmetric matter in the free QGP limit.
The resulting chemical freeze-out line, shown by the blue dotted line in Fig.~\ref{fig:frzCP}, is very close to the heavy-ion line obtained under $n_S = 0$, $n_Q/n_B = 0.4$ conditions.
Setting $\mu_S = \mu_B / 3$ is sometimes used to approximate the strangeness-neutral equation of state in heavy-ion collisions~(see e.g. \cite{Gunkel:2021oya}), and our HRG model calculation presented here indicates that such an approximation is reasonably accurate.

We now discuss the various CP estimates and their comparison with the F-O bound in more detail.
The LQCD-YLE estimate from Parma-Bielefeld group~\cite{Clarke:2024ugt} is based on extracting Yang-Lee edge singularities from imaginary $\mu_B$ simulations via multipoint Pad\'e approach. 
The extraction is performed at temperatures $T \geq 120$~MeV, and extrapolation toward a CP at a lower temperature is performed based on the expected critical scaling behavior of YLE~\cite{Stephanov:2006dn}.
The extracted CP location is $T_{CP} = 105^{+8}_{-18}$ MeV and $\mu_{B, \ CP} = 422^{+80}_{-35}$~MeV, which is significantly below the F-O curve as shown in Fig.~\ref{fig:frzCP}.
However, the primary analysis in Ref.~\cite{Clarke:2024ugt} was performed on coarse lattices~($N_{\tau} = 6$), and the importance of taking the continuum limit was emphasized. 
Possible systematic uncertainties in determining YLE singularities and the extrapolation from higher temperatures down to the CP needs also be controlled. Analyses of $N_\tau = 8$ data sets using Pad\'e method in~\cite{Clarke:2024ugt} and using various resummations in Ref.~\cite{Basar:2023nkp} indicate larger values of $\mu_{B, \ CP} \approx 560$--$580$~MeV, and the temperature of $T_{CP} \approx 100$~MeV, with sizable errors.
These estimates are closer to the F-O curve and indicate the importance of continuum extrapolation, along with that of the control over systematic uncertainties in determining YLE singularities and the extrapolation from high temperatures down to the CP.

A different lattice-based estimation of the CP has recently been presented in Ref.~\cite{Shah:2024img}.
It is based on the appearance of crossings of entropy density contours, which have been extrapolated from $\mu_B = 0$ based on continuum-extrapolated lattice data on entropy density and baryon number susceptibility at vanishing baryon density.
This LQCD-S estimate, obtained in Ref.~\cite{Shah:2024img} on a $\mathcal{O}(\mu_B^2)$ level places the CP at $T_{CP} = 114.3 \pm 6.9$~MeV, $\mu_{B, \ CP} = 602.1 \pm 62.1$~MeV, which is above the F-O line presented here.

The LQCD-S results are consistent with the predictions of functional approaches, based on functional renormalization group~(fRG)~\cite{Fu:2019hdw}, or Dyson-Schwinger equations~(DSE)~\cite{Gao:2020fbl,Gunkel:2021oya}, depicted in Fig.~\ref{fig:frzCP} by different symbols and which all also lie above the F-O curve.\footnote{Note that the DSE estimate from Ref.~\cite{Gunkel:2021oya} is obtained for $\mu_Q=0$, $\mu_S=\mu_B/3$ case instead of $\mu_Q = \mu_S = 0$. 
This CP estimate should therefore be contrasted with the blue dotted line in Fig.~\ref{fig:frzCP}, and it is positioned higher than this line.} The question of systematic error in functional approaches is important and has been estimated in fRG~\cite{Fu:2019hdw} to give the range $(135,480)~\text{MeV} \lessapprox (T_{CP}, \mu_{B, \ CP}) \lessapprox (103,660)~\text{MeV}$ for the possible location of the CP.
Finally, the estimates from Ref.~\cite{Hippert:2023bel} are based on a holographic black hole model fixed to reproduce lattice QCD thermodynamics at $\mu_B = 0$.
These estimates place the CP very close to the F-O curve.
 
One can see that the various estimates presented here place the CP in a similar region of the phase diagram, and some of them are very close to the F-O line.
It should be noted that the HRG model is expected to break down as one approaches the CP of the QCD phase transition, as the model has no phase transition or partonic degrees of freedom.
Therefore, one could argue that the CP should not only lie above F-O curve, but also be some distance away from the F-O curve. 
It is not trivial to estimate precisely how far away from the phase transition the HRG model description of hadron yields still applies, but this argument does indicate that the obtained F-O line favors LQCD-S/DSE/fRG estimates that are $10$--$15$ MeV in temperature above the F-O line somewhat more than LQCD-YLE and BHE ones.

To estimate the relevant collision energy range where the F-O lies the closest to the CP estimates, we utilize the following parametrization of the collision energy dependence of $\mu_B$~\cite{Vovchenko:2015idt}:
\eq{ \label{eq:paramMU}
    \mu_{B} \cong  \frac{c}{1 + d\sNN}
}
with parameter values\footnote{The parameter errors are immaterial for the discussion in the present paper.} of $c = 1.477$~GeV and $d = 0.343$~ GeV$^{-1}$. 
By utilizing Eq.~\eqref{eq:paramMU}~(see black points in Fig.~\ref{fig:frzCP}), one can estimate the collision energy range of $\sNN = 3.5$--$6$~GeV as most promising in the search for the CP, which can be explored with the measurements of proton cumulants within RHIC fixed target program and the future CBM experiment at FAIR~\cite{CBM:2016kpk,Almaalol:2022xwv}. 
The lower energies, $\sNN \lessapprox 3.5$~GeV, on the other hand, can be utilized to look for the signs of a first-order phase transition.

\section{Summary}
\label{sec:summary}

In the present work we pointed out that the chemical freeze-out (F-O) line in heavy-ion collisions provides a lower bound in temperature on the possible location of the QCD critical point (CP). While the F-O line is subject to uncertainties at large baryochemical potential $\mu_B$, we observe that a constant energy per particle line $\varepsilon/n = 0.951$~GeV evaluated within the ideal HRG model provides a reasonable estimate for the lower bound on the chemical F-O temperature at $\mu_B > 400$~MeV and thus also sets the lower bound on QCD CP. 

Here we evaluated in detail the effect of strangeness neutrality condition and found that its removal leads to a downward shift of the F-O temperature at fixed $\mu_B$. 
Comparing the F-O line under $\mu_S = \mu_Q = 0$ conditions to recent CP location estimates, we find that most of the recent CP estimates, based either on effective QCD approaches or extrapolation of lattice QCD data, place the CP slightly above the F-O curve, suggesting that heavy-ion collisions at $\sNN = 3.5$--$6$~GeV may be most sensitive to the CP. 
We also observe that a CP estimate based on Yang-Lee edge singularities extracted from $N_\tau = 6$ lattice QCD simulations lies significantly below the F-O line, likely indicating the importance of achieving the continuum limit in that approach.

We do point out that our conclusions are based on the validity of an equilibrated hadron resonance gas picture at the chemical freeze-out stage of heavy-ion collision, in particular at intermediate collision energies that probe the baryon-rich region of the QCD phase diagram where the CP may be located.
Although there is significant evidence from the measured hadron abundances that this is the case, one has to keep this assumption in mind when considering the F-O bound on CP presented in this paper.
On the other hand, if convincing evidence does emerge that the CP is located below the phenomenological freeze-out curve, the concept of chemical freeze-out may need to be revisited.

{\bf Acknowledgements.}
~
A.L. thanks Prof. S. Vilchinskii for his useful advice and assistance, which played a major role in writing this paper. 
V.V. thanks S. Bors\'anyi for fruitful discussions on the strangeness neutrality condition.
M.I.G. is thankful for the support from the Simons  Foundation.

\appendix
\section{Freeze-out line in non-ideal HRG models} 
\label{sec:A}

Here we analyze the possible influence of baryonic interactions on the chemical F-O line by evaluating the $\varepsilon / n = \const$ line in EV-HRG and QvdW-HRG models.

The pressure in the QvdW-HRG model reads
\eq{ \label{Pvdw1}
&    p(T, \mu) 
=P_{M}(T, \mu)
 + P_{B}(T, \mu)
    + P_{\bar{B}}(T, \mu)
}
with
\eq{ \label{PM}
&    P_{M}(T, \mu) =
\sum_{i \in M} p_{i}^{\rm id}(T, \mu_{i})~\textbf{,}\\
 \label{PB}
&    P_{B}(T, \mu) = 
\sum_{i \in B} p_{i}^{\rm id}(T, \mu_{i}^{B*}) - an_B^2~\textbf{,}\\ 
\label{PB_}
&    P_{\bar{B}}(T, \mu) 
= \sum_{i \in \bar{B}} p_{i}^{\rm id}(T, \mu_{i}^{\bar{B}*}) - an_{\bar{B}}^2~\textbf{,}
}
where $\mu_{i}^{B(\bar{B})*} = \mu_i - bP_{B(\bar{B})} - 
abn_{B(\bar{B})}^2 + 2an_{B(\bar{B})}$, $n_{B(\bar{B})}$ is the number density of baryons (antibaryons), which is determined from the following expression:
\eq{ \label{nBB_}
    n_{B(\bar{B})} = (1 - bn_{B(\bar{B})}) \sum\limits_{i \in B(\bar{B})} n_{i}^{\rm id}(T, \mu_{i}^{B(\bar{B})*})
}
and $a$, $b$ are the constants of the van der Waals quantum equation of state.

The total particle number density in the QvdW-HRG model is calculated as
\eq{
n= n_M~+~n_B~+ n_{\bar{B}}~\textbf{,}
}
where baryon and antibaryon densities are given by Eq.~(\ref{nBB_}) and meson number density  by the ideal gas expression:
\eq{
n_M =\sum_{i \in M}n_i^{\rm id}(T,\mu_i)~\textbf{.}
}
The total energy density can be calculated as
\eq{ \label{EQvdW}
&   \varepsilon(T, \mu)=
    T\left(\frac{\partial p}{\partial T}\right)_{\mu}
     + \mu_{B}\left(\frac{\partial p}{\partial \mu_{B}}\right)_{T,\mu_Q,\mu_S} \\ 
     & + \mu_{Q}\left(\frac{\partial p}{\partial \mu_{Q}}\right)_{T,\mu_B,\mu_S}
     + \mu_{S}\left(\frac{\partial p}{\partial \mu_{S}}\right)_{T,\mu_B,\mu_Q} ~ - ~ p~\textbf{.} \nonumber 
}

\begin{figure*}
    \includegraphics[scale=0.5]{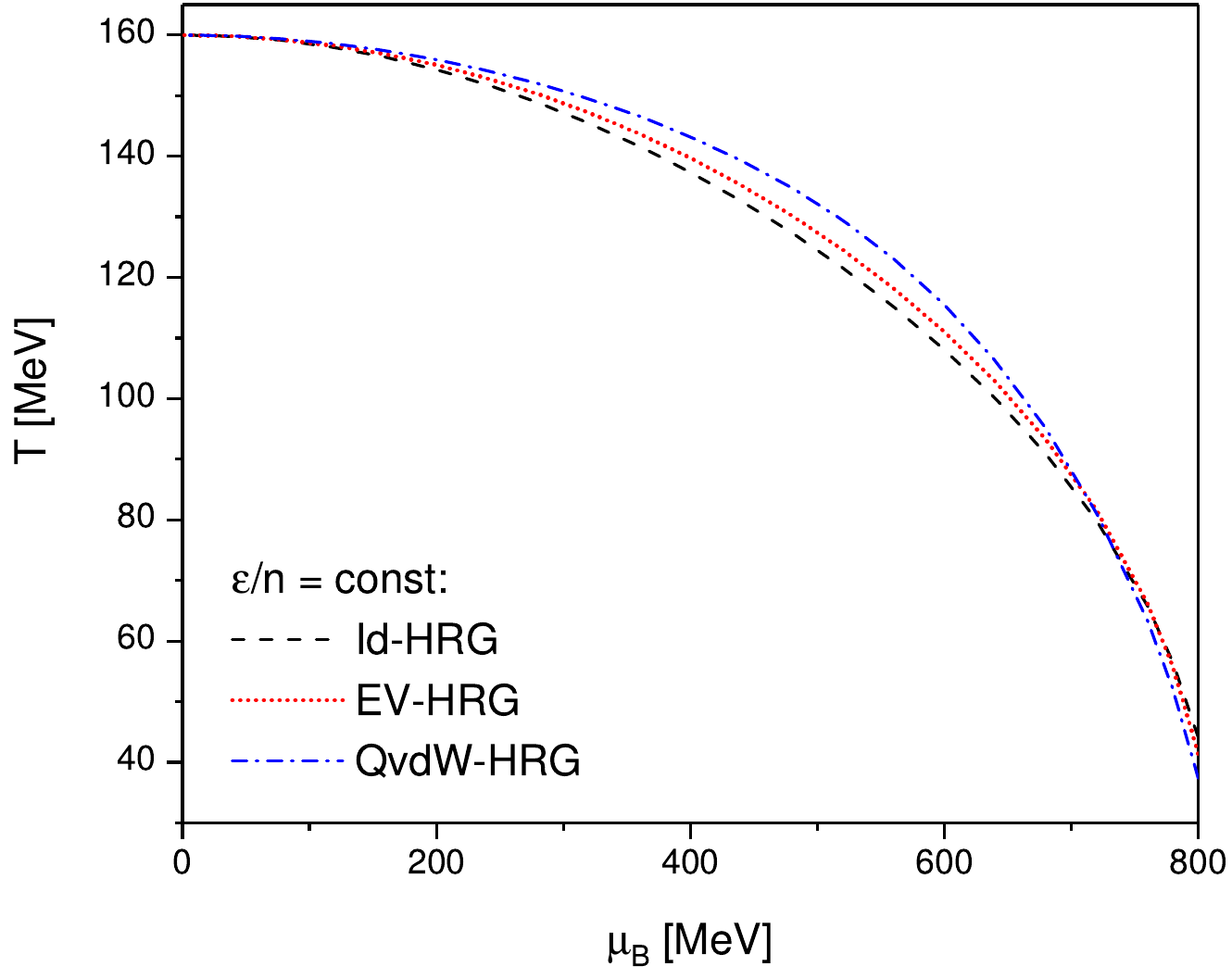}
    \caption{
    Constant energy per particle lines corresponding to chemical F-O in heavy-ion collisions calculated within three versions of the HRG model: Id-HRG~(dashed black line), EV-HRG~(dotted red line), and QvdW-HRG~(dash-dotted blue line).
    }
    \label{fig:porivHRG}
\end{figure*}

It can be seen from  Eqs. (\ref{Pvdw1})-(\ref{nBB_}) that at $a = 0$ and $b = 0$, the system reduces to the Id-HRG. The special case where $a = 0$ but $b$ remains nonzero is referred to as an excluded volume hadron resonance gas (EV-HRG). This scenario represents a system where only repulsive interactions between (anti)baryons are present. 

In this work, we use the values $a = 329 \ \rm MeV \ fm^3$, $b = 3.42 \ \rm fm^3$ for calculations within QvdW-HRG model, and $b = 1 \ \rm fm^3$ within the EV-HRG model. The former parameter set corresponds to the ground state of nuclear matter at temperature $T = 0$~\cite{Vovchenko:2015vxa}, while the latter provides a good description of baryon number susceptibilities from lattice QCD at $\mu_B = 0$ and $T < 160$~MeV~\cite{Vovchenko:2017gkg}.

The results of the calculations are presented in Fig.~\ref{fig:porivHRG}. The lines were plotted under the condition $\varepsilon/n = \rm const$, with the value of the constant chosen differently for each model to ensure $T(\mu_B=0)=160 \ \rm MeV$. Namely,
\begin{itemize}
\setlength\itemsep{0.1em}
    \item Id-HRG: 0.951 GeV,
    \item EV-HRG: 0.946 GeV,
    \item QvdW-HRG: 0.942 GeV.
\end{itemize}

One observes that the three lines do not differ much. The inclusion of interactions generally shifts the temperature slightly upward (see also Ref.~\cite{Poberezhnyuk:2019pxs}). Thus one can regard the Id-HRG model calculation as providing a lower bound on the F-O temperature at fixed $\mu_B$.

\bibliography{references}

\end{document}